%% file: main.tex
\documentclass[a4paper,superscriptaddress, amsfonts, amssymb, amsmath, reprint, showkeys, dvipsnames, nofootinbib, twoside, aps,prl,longbibliography,nobibnotes]{revtex4-1}

\usepackage[english]{babel}
\usepackage[utf8]{inputenc}
\usepackage[colorinlistoftodos, color=green!40, prependcaption]{todonotes}

\usepackage{amsthm}
\usepackage{mathtools}
\usepackage{graphicx}
\usepackage{adjustbox}
\usepackage{placeins}
\usepackage[T1]{fontenc}
\usepackage{lipsum}
\usepackage{csquotes}
\usepackage{xcolor}
\usepackage{soul}

\usepackage{units}
\usepackage[colorlinks = true, linkcolor = blue,
            urlcolor  = blue,citecolor = blue,
            anchorcolor = blue]{hyperref}
\usepackage[noabbrev]{cleveref}

\begin{document}

\title{Demonstration of Flying-Focus-Enhanced Relativistic Thomson Scattering}



\include{authors}

\date{\today} 
\begin{abstract}


We report the experimental demonstration of a spatiotemporally engineered ``flying-focus'' laser pulse for enhanced X-ray generation via relativistic Thomson scattering. 
By combining longitudinal chromatic aberration, angular dispersion, and group delay dispersion, we create a relativistically intense laser pulse whose focal point propagates with an independently tunable velocity vector, enabling control of both its speed and propagation angle. 
Precise tuning of the focal-point trajectory is used to match the motion of a near-counterpropagating ultrarelativistic electron bunch generated by a laser wakefield accelerator. 
This spatiotemporal velocity matching extends the interaction length while mitigating nonlinear scattering effects, resulting in an enhanced X-ray yield. 
Simulations indicate that this approach can increase the spectral density and brightness of Thomson-scattered X-ray beams by orders of magnitude relative to equivalent focusing without spatiotemporal control. 
The technique provides a new degree of freedom for high-intensity laser-matter interactions, allowing dynamic intensity structures with programmable two-dimensional motion and opening a route to enhanced laser-particle interactions and next-generation ultrabright radiation sources.
\end{abstract}

\maketitle

\section*{Introduction}

An emerging class of spatiotemporal pulse-shaping techniques offers new degrees of freedom for tailoring the propagation of intense laser fields in space and time \cite{Piccardo_Optica2025}.
Among these, the ``Flying Focus'' concept \cite{Sainte-Marie2017Optica,Froula2018NP} provides a particularly promising approach.
By combining longitudinal chromatic aberration (LCA) and group delay dispersion (GDD), different frequency components focus at different times and positions \cite{Froula2018NP,Jolly_OptExp2020}.
The resulting focal point can be made to move along a programmable trajectory at a tunable velocity.
This \emph{chromatic} flying focus sustains the peak intensity over extended distances and along arbitrary paths, offering a route to overcome diffraction and synchronisation constraints that limit many high-intensity laser-plasma interactions and fundamental physics \cite{Palastro_PRL2020,Caizergues2020NP,Formanek2022PRA,Miller_SciRep2023,Jin2023PRA,Gong_PRL2024,Formanek2024PRD}.

\begin{figure*}
    \centering
    \includegraphics[width=17.4cm]{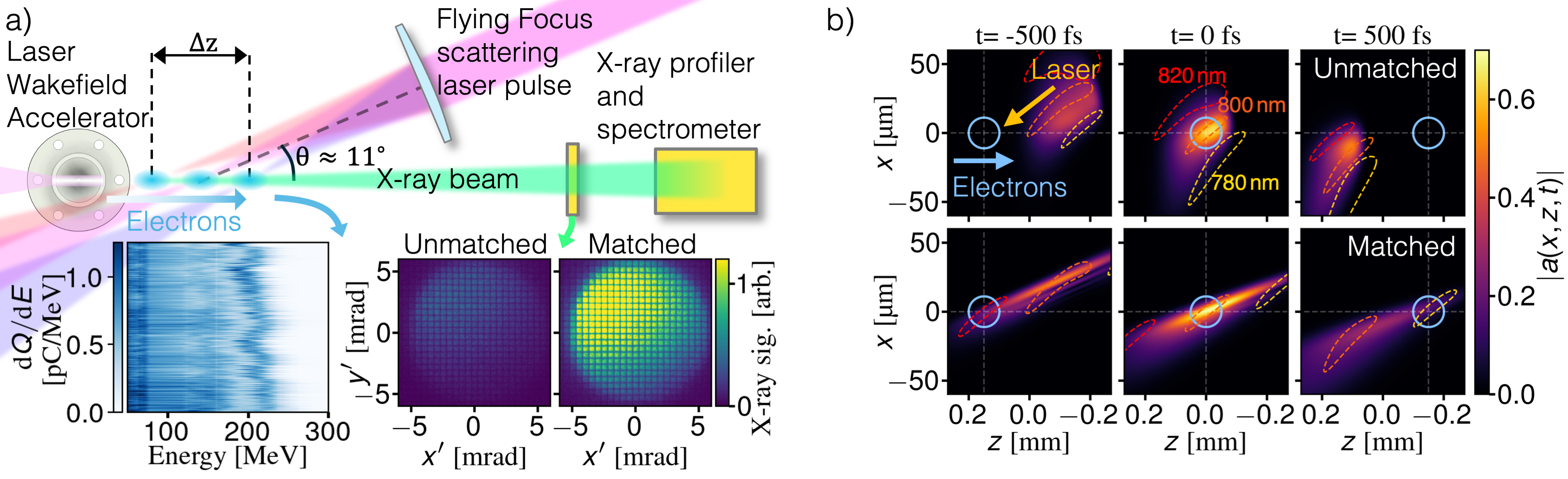}
    \caption{\textbf{A two-dimensional flying focus enables an extended overlap between the focal point and electron bunch.}
    \textbf{a)} Illustration of the experimental geometry with example electron spectra from 200 consecutive shots (inset left) produced by the laser wakefield accelerator.
    A secondary laser pulse was modified to generate an angled chromatic flying focus that tracked the location of the near-counterpropagating electron bunch.
    When the electron bunch and flying-focus trajectories were matched, the Thomson scattering signal was maximised (see X-ray profiles inset right).
    \textbf{b)} Calculations of the laser field amplitude in the focal region during the interaction with the electron bunch. 
    The top row shows the chromatic focus with zero GDD, with the electron bunch (position indicated by the blue circles) crossing through the laser field.
    The bottom row shows the same spatial-spectral configuration but with matched GDD, such that the focus overlaps with the electron trajectory at all times, extending the interaction length.
    The locations of the spectral components centred at wavelengths of \unit[780]{nm}, \unit[800]{nm}, and \unit[820]{nm} are indicated by the dashed contours.
    }
    \label{fig:conceptual}
\end{figure*}

The flying focus has been proposed as a method for enhancing relativistic Thomson scattering \cite{McAnespie_PMB2022,Ramsey2022PRE,Ye_AIP2023}.
In Thomson scattering, a relativistic bunch of electrons, with energies $E_e = \gamma m_e c^2$, collides with an intense laser pulse with initial photon energies of $E_i = \hbar\omega_i$, causing emission of photons at much higher frequencies $E_f \propto \gamma^2 E_i$, extending into the X-ray range or beyond.
The total scattered energy for a given electron bunch scales as $\mathcal{E}_{S} \propto  a_0^2 \tau$, where $a_0$ is the peak normalised vector potential of the laser pulse and $\tau$ is the interaction duration.
At relativistic intensities, i.e., $a_0\gtrsim1$, electron motion in the laser field becomes nonlinear, leading to the emission of harmonics and an increase in angular divergence of the emitted X-rays \cite{Esarey1993PRE,Yan2017NatPhoton}.
This results in a broadened spectrum and emission cone, limiting the spectral brightness and adversely impacting the utility of the X-rays for applications.

To maximise the spectral density and brightness, it is advantageous to operate at moderate intensity over a longer interaction length. 
This can provide the same total scattered energy while avoiding the adverse effects of nonlinear electron motion.
Meanwhile, achieving high efficiency, i.e., maximising scattered energy for a given laser pulse energy, also requires tight focusing of the pulse.
This creates a conflict as diffraction inherently limits the overlap between the focused laser and the electron bunch; i.e., tight focusing and long interaction length are in direct opposition unless externally guided \cite{Rykovanov2014JPB}.
The flying focus enables extended interaction lengths by matching the motion of the narrow laser focus to an electron bunch, providing a pathway to enhancing the efficiency of X-ray generation for a given laser system.

Previous work has applied the chromatic flying-focus technique in experimental demonstrations of controlled plasma generation \cite{Turnbull2018PRL,Froula2018NP} and soft X-ray laser amplification \cite{Kabacinski_NatPhoton2023}, making use of diffractive and refractive optics to create the spatiotemporal coupling. 
More geometric implementations, using novel reflective optics such as axiparabolas or echelons, promise robust operation at high power and ultrashort pulse durations \cite{Palastro_PRL2020}, but with a reduced range of focal velocities compared to the chromatic approach.
These more geometric approaches have recently been used to demonstrate the flying focus at relativistic intensities to drive a plasma wakefield \cite{liberman2025NatComms} and for the acceleration of electrons \cite{Liberman2025arxiv2,Arrowsmith2026NatPhys}.

Here we report on the experimental demonstration of a high-intensity chromatic flying focus that enhances the brightness of X-rays produced from Thomson scattering.
LCA, introduced by the final focusing lens, was coupled with angular dispersion and GDD established in the grating compressor.
This resulted in a spatiotemporally engineered laser pulse that traced an elongated focal region at a tunable angle and velocity, providing two-dimensional (2D) control \cite{Cao2026PRA}.
The angle and velocity of the 2D flying focus were precisely tuned to match the trajectory of a counterpropagating ultrarelativistic electron bunch, generated via laser wakefield acceleration (LWFA).
The flying focus increased the number of photons detected in the \unit[0.1--1.0]{MeV} range incident on the detector by more than a factor of two compared to the X-ray spectrum computed for equivalent focusing without spatiotemporal control.

A benefit of this novel configuration is that it enables interactions at a finite angle, protecting the laser chain from backpropagating laser light and providing a clear path for radiation to propagate to detectors and applications.
By using a 2D flying focus, we are able to extend the interaction length without losing overlap due to the angle between the electron bunch and the scattering laser propagation axis.
The properties of the measured X-ray radiation and its dependence on GDD were in good agreement with modelling.
Through this interaction, the electron bunch acted as a relativistic probe of the local field structure and its evolution, verifying the programmed properties of the flying focus.
A 2D flying focus as demonstrated here could also benefit other high-intensity applications, including ion acceleration, electron acceleration, and terahertz generation.



\section*{Results}

The experiment was performed at the Lawrence Berkeley National Laboratory (LBNL) Laser Accelerator (BELLA), using the Hundred Terawatt Thomson facility (HTT) \cite{Tsai2026SciRep}.
This laser system generates two separate laser pulses (both amplified from a common seed pulse) which propagate through individual pulse compressors, allowing the GDD of the two pulses to be tuned independently.
One laser pulse was used to generate an electron beam, and the other provided the colliding pulse for the Thomson scattering interaction.
A schematic of the experimental geometry is shown in \cref{fig:conceptual}a.


The electron bunches had a broadband spectrum with a 95th percentile energy of \unit[$(210\pm20)$]{MeV}, an average charge of \unit[$(140\pm40)$]{pC} for energies above \unit[25]{MeV}, and an average rms divergence of $\unit[(5.0\pm0.5)]{mrad} \times \unit[(2.3\pm0.2)]{mrad}$, in the horizontal and vertical directions, respectively.
The \unit[0.5]{J} scattering laser was focused by a fused silica plano-convex lens with a focal length of \unit[$f_0 = 254$]{mm} at \unit[$\lambda_0=800$]{nm} to a FWHM spot size of \unit[$(5.5 \pm 0.3)$]{{\textmu}m} (when the bandwidth was reduced to \unit[3]{nm}).
The LCA introduced by this optic resulted in a wavelength-dependent focal plane $f_{\lambda} = f_0 + \left(\mathrm{d}f/\mathrm{d}\lambda \right)\left(\lambda - \lambda_0\right)$, with a chromaticity of \unit[$\mathrm{d}f/\mathrm{d}\lambda = 9.9$]{{\textmu}m/nm}, resulting in a FWHM focal range of \unit[$\Delta f \approx210$]{{\textmu}m} for a laser bandwidth of \unit[$\Delta \lambda \approx21$]{nm}.
The interaction angle between the LWFA electron bunch and the scattering laser pulse was determined to be $(168.9\pm0.8)^{\circ}$, i.e. $\theta = (11.1\pm0.8)^{\circ}$ from counterpropagating.
An angular dispersion of \unit[$(6 \pm 1)$]{{\textmu}rad/nm} was introduced by rotating a grating in the scatter beam compressor by $0.155^{\circ}$.
This parallelised the scatter line focus to the electron bunch axis to within $(2\pm2)^{\circ}$, corresponding to a maximum misalignment of approximately \unit[7]{{\textmu}m} over the length of the \unit[$210$]{{\textmu}m} focal range.
The focal velocity was then set by changing the grating separation to tune the GDD while the relative timing was set with a motorised delay stage.

\subsection*{Matching focal velocity to the electron bunch} 

\begin{figure}
    \centering
    \includegraphics[width=8.45cm]{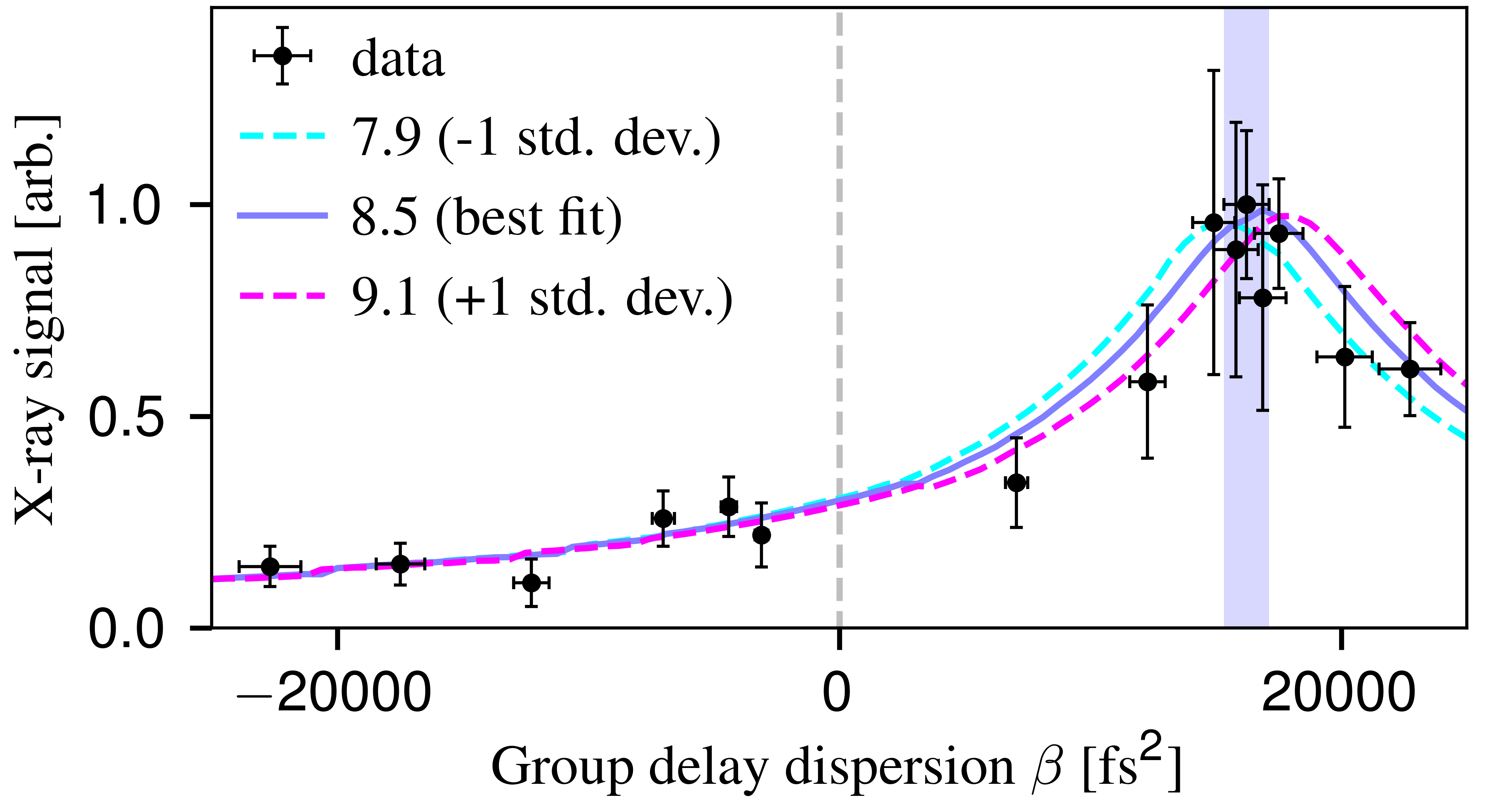}
    \caption{\textbf{Optimisation of the flying-focus velocity.}
    The measured X-ray signal from Thomson scattering (in arbitrary units) as a function of GDD ($\beta$) for the scattering laser, showing that the X-ray signal is maximised when the trajectory of the flying focus matches that of the electron bunch. 
    Vertical error bars are the rms shot-to-shot fluctuations, while the horizontal error bars represent calibration uncertainties.
    The zero-GDD point ($\beta=0$, grey dashed) corresponds to optimal compression.
    The highest X-ray signal was observed for the matched flying focus at \unit[$\beta=(16200\pm900)$]{fs$^2$} (blue shaded region).
    The model predictions, based on scattering calculations of test particles propagating through the simulated field, are shown for the best fitting flying-focus angle of $8.5^{\circ}$ (solid line), and for the angles bounding the 68\% confidence interval (dashed line).
    }
    \label{fig:xray_vs_chirp}
\end{figure}

The X-ray counts as a function of scatter pulse GDD, averaged over 10--100 repeat shots at the optimal scattering laser alignment and delay relative to the electron bunch, are shown in \cref{fig:xray_vs_chirp}.
The X-ray signal shows a maximum for a GDD of \unit[$\beta=(16200\pm900)$]{fs$^2$}, indicating that the trajectory of the flying focus was closely matched to that of the ultrarelativistic electron bunch for this value.
The X-ray signal dropped considerably as the GDD was changed from the optimum, due to the reduced overlap between the electron bunch and scattering pulse.

Numerical simulations of the scattering field near the focus were consistent with the observed X-ray signal as a function of GDD.
The expected X-ray yield was calculated from the path integral through the simulated scattering field, $\int a(\mathbf{x}(t))^2 \mathrm{d}{t}$, for an ultrarelativistic point-particle with time-dependent position $\mathbf{x}(t)$ and a spatial offset, $\mathbf{x}_0 = \mathbf{x}(t=0)$, chosen to maximise the integral.
The integrals were performed for fields with varying levels of angular dispersion, resulting in different angles of the  chromatic line focus, and for electron trajectories propagating at an angle of $11.1^{\circ}$.
The amplitude of each model curve was scaled to best fit the experimental data for \unit[$\beta < -5000$]{fs$^2$}, where the yield was relatively insensitive to changes in the flying-focus and electron trajectories.
Monte Carlo error propagation (see methods for details) was used to determine the mean best-fitting angle and its standard deviation.
The model evaluated at the mean best-fitting angle is shown as the solid line in \cref{fig:xray_vs_chirp}, while the models evaluated at $\pm1$ standard deviation angles of best fit from this ensemble are shown as dashed lines.

The closest agreement between the simulated and experimental X-ray yields was obtained for a chromatic line focus angle of $(8.5\pm0.6)^{\circ}$ from counterpropagating.
This best-fit value should be interpreted as an effective line-focus angle in the optical model, which can differ from the design value due to alignment and calibration uncertainties.
The overall yield is relatively insensitive to this residual angular mismatch because the Rayleigh range of the laser pulse is large compared to the associated spatial offset; the primary effect is a shift in the value of $\beta$ required to match the flying-focus velocity to the electron trajectory.
When matched, as shown in \cref{fig:conceptual}b, the flying-focus pulse had a FWHM duration of \unit[$\tau_L = 1.0$]{ps}, and a peak normalised vector potential of $a_0 \approx 0.7$.
As a consequence of the applied LCA and angular dispersion, the peak $a_0$ only varied by $\approx 10$\% over the full range of GDD values shown in \cref{fig:xray_vs_chirp}.

\begin{figure}
    \centering
    \includegraphics[width=8.45cm]{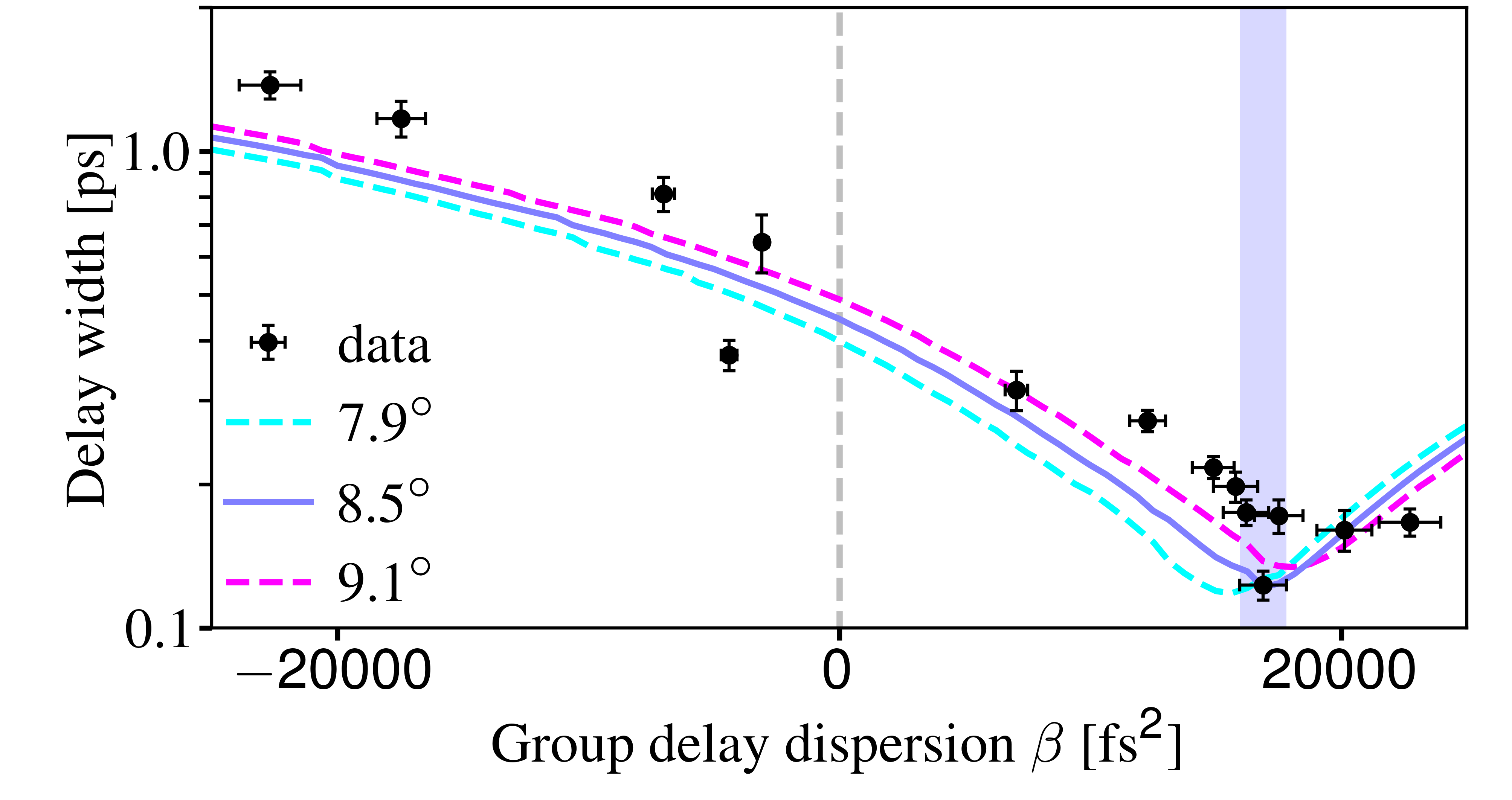}
    \caption{\textbf{Sensitivity of the X-ray signal to timing of the flying focus.}
    The rms temporal synchronisation sensitivity as a function of GDD.
    The data points show the rms of a Gaussian function fitted to the X-ray signals as functions of relative delay between the LWFA drive laser and the scattering laser.
    Vertical error bars are the standard deviation estimates from the parameter fitting, while the horizontal error bars represent calibration uncertainties.
    The predicted synchronisation sensitivity (from the optical model) is plotted for the same focus angles as \cref{fig:xray_vs_chirp}.
    The point of highest timing sensitivity occurred at \unit[$\beta=(16900\pm900)$]{fs$^2$} (blue shaded region).
    }
    \label{fig:delay_scan}
\end{figure}

Further evidence of precise matching of the flying focus to the electron trajectory was seen in the synchronisation sensitivity between the LWFA drive and Thomson scattering laser pulses.
For the matched GDD, the X-ray signal was highly sensitive to timing, whereas for highly unmatched cases, the electron bunch interacted with at least some part of the pulse over a wide range of delays, as illustrated in \cref{fig:conceptual}b.
To quantify the synchronisation sensitivity, the relative delay between the scattering laser and the LWFA drive laser was scanned for each value of GDD.
A Gaussian function was fitted to the resulting X-ray signal as a function of delay, with the rms width taken as the synchronisation sensitivity. 
The results are shown in \cref{fig:delay_scan}, along with predictions from the calculated optical fields.
The highest temporal sensitivity was seen for \unit[$\beta=(16900\pm900)$]{fs$^2$}, consistent with the numerical model, a difference of \unit[700]{fs$^2$} compared to the value that produced the maximum X-ray signal.

\subsection*{Flying-focus-enhanced X-ray generation}

\begin{figure}
    \centering
    \includegraphics[width=8.45cm]{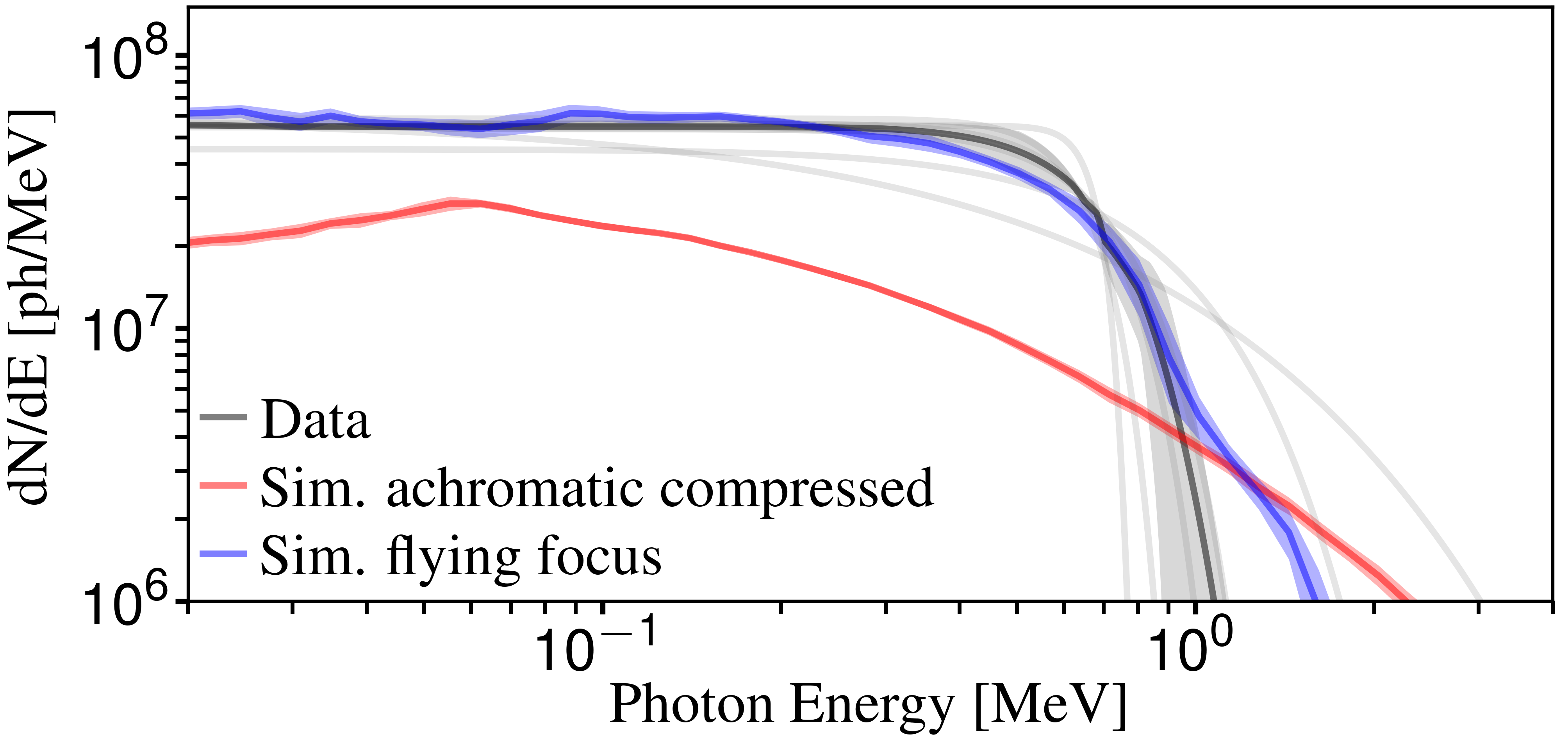}
    \caption{\textbf{Measured and simulated X-ray spectra incident on the detector}.
    Measured X-ray spectra from Thomson scattering (median shown in black, with individual shots in grey and the median absolute deviation in the shaded region) in good agreement with the simulated emission spectrum (from Ptarmigan) for a matched flying focus (blue), scaled by the measured photon yield.
    The simulated spectrum from a fully compressed laser pulse without spatiotemporal control extends to higher photon energies with a synchrotron-like spectral shape (red), but at lower spectral density below \unit[1]{MeV} incident on the detector than in the matched flying-focus case.}
    \label{fig:xray_spectrum}
\end{figure}

The measured X-ray spectrum for the matched flying focus is shown in \cref{fig:xray_spectrum} (the retrieval procedure is described in Methods). 
The interaction generated an average of $(3.6 \pm 1.0) \times 10^7$ (standard deviation) photons above \unit[20]{keV} incident on the detector, with a maximum of $(5.8 \pm 2.0) \times 10^7$ (systematic error) on the single brightest shot.
The photon beam had an rms divergence of \unit[$(5.7\pm1.1)$]{mrad} (standard deviation).
The spectrum was broadband due to the relatively large energy spread and divergence of the electron bunch during the scattering interaction.

The properties of the X-ray pulse were simulated using Ptarmigan \cite{Blackburn_PoP2023}, which calculates the trajectories of charged particles in a specified electromagnetic field, including quantum radiation reaction, and the production of secondary particles, such as X-rays and electron-positron pairs.
The resulting X-ray beam was then angularly filtered to the central \unit[$\pm 3.2$]{mrad} to match the angular range of the experimental detector.
The case of the matched flying focus predicts a spectrum that closely agrees with the shape of the retrieved spectrum in the experiment, as shown in \cref{fig:xray_spectrum}, when scaled to match the total detected energy.
The qualitative agreement of the spectral shape indicates that the interaction with the matched flying-focus pulse was close to linear.
We also show the calculated X-ray spectrum (incident on the detector) for the case of a fully compressed scattering pulse without spatiotemporal control.
In this case, $a_0=5.2$, and so the spectral shape exhibits significant spectral broadening due to nonlinear electron motion.
Also, the angular divergence of the X-ray beam increases from \unit[6]{mrad} for the flying-focus case to \unit[16]{mrad} in the laser polarisation axis for the fully compressed case.
The combination of these factors results in more than twice the number of photons between \unit[0.1--1.0]{MeV} incident on our detector for flying-focus-enhanced Thomson scattering.
This enhancement of the near on-axis spectrum is obtained even though scattering with the compressed pulse would generate more than twice the total radiated energy than the flying-focus case, with a comparable energy in the same angularly-integrated spectral range.

The generated photon yield also depends on the transverse size of the electron bunch.
At the interaction plane, \unit[0.8]{mm} downstream of the plasma exit, the electron bunch is expected to be comparable in transverse size to the laser focus, such that electrons experience a range of field strengths depending on their transverse position. 
Combining the optical model of the flying focus with the measured average number of photons emitted per electron, we infer an rms electron bunch size of \unit[8]{{\textmu}m} at the interaction plane.
With this electron bunch size, the yield is approximately 5\% of the maximum photon number expected for a zero-size electron bunch.

\section*{Discussion}

\begin{figure}
    \centering
    \includegraphics[width=8.45cm]{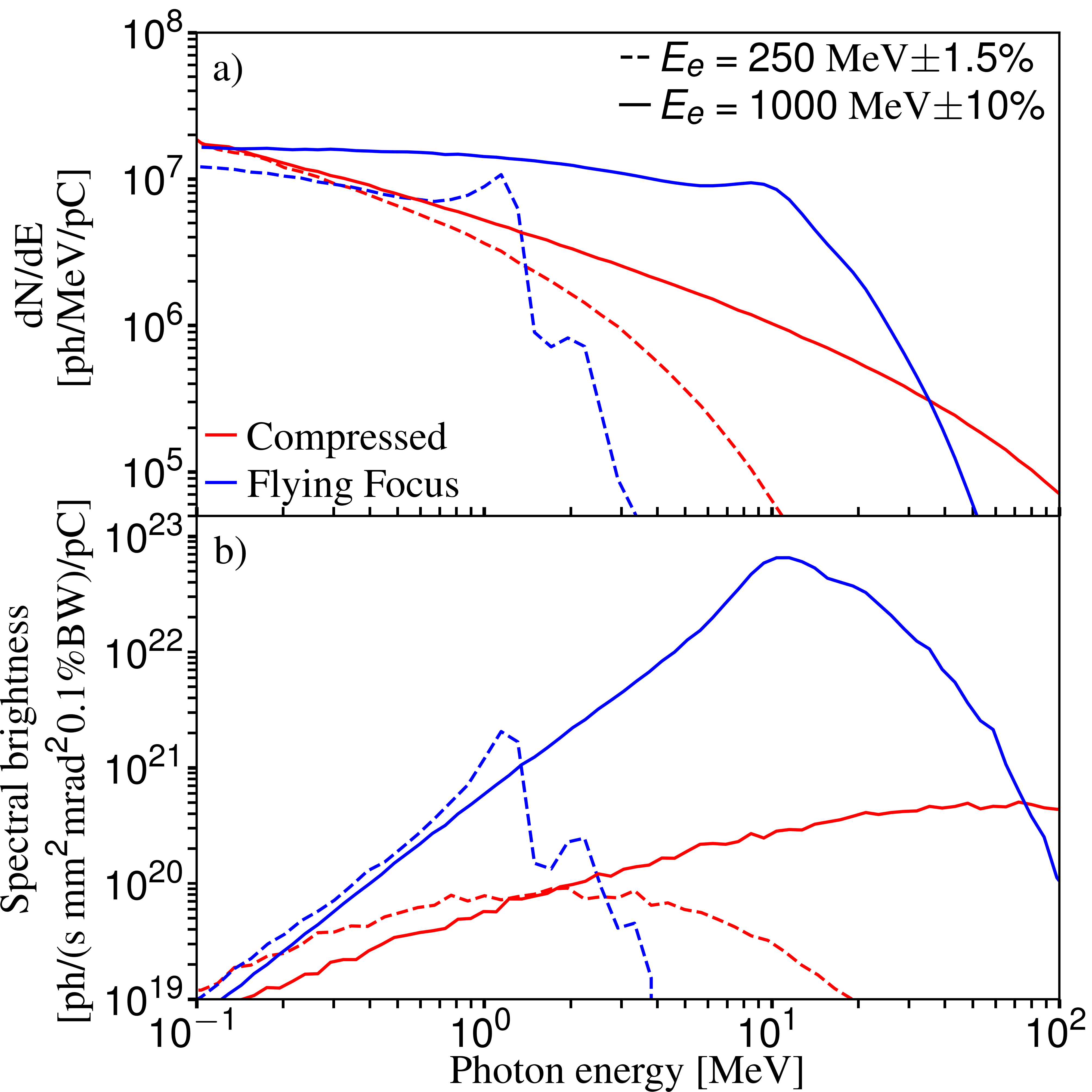}
    \caption{\textbf{Flying-focus enhancement of X-ray spectrum and brightness.}
    \textbf{a)} The angularly integrated X-ray spectra from Thomson scattering simulations using Ptarmigan \cite{Blackburn_PoP2023} for two example configurations.
    In one example (dashed lines), a \unit[$E_e=250(\pm1.5\%)$]{MeV} electron bunch with \unit[1]{mrad} rms divergence collides with a \unit[0.5]{J} laser pulse at a collision angle of $10^{\circ}$.
    Results are shown both for a matched flying focus ($\tau_L=\unit[1.3]{ps}$, $a_0=0.6$, blue) and equivalent conventional focusing with a fully compressed pulse ($\tau_L=\unit[35]{fs}$, $a_0= 5.5$, red), where $\tau_L$ is the laser pulse FWHM duration.
    In the second example (solid lines), a \unit[$E_e=1(\pm10\%)$]{GeV} electron bunch with \unit[0.25]{mrad} rms divergence collides with a \unit[10]{J} laser pulse at a collision angle of $5^{\circ}$.
    In this case, the fully compressed pulse reaches $a_0 \approx 16.6$ (red), while for the matched flying focus $a_0 \approx 1.0$ ($\tau_L=\unit[6.4]{ps}$, blue).
    \textbf{b)} Spectral brightness calculated from the simulated X-ray spectra and their associated energy-dependent divergence, assuming a \unit[30]{fs} duration electron bunch, and a source size of \unit[5]{{\textmu}m} (dashed) and \unit[7]{{\textmu}m} (solid).
    }
    \label{fig:future_xray_spectrum}
\end{figure}

Our experiment demonstrates the ability to generate a chromatic flying focus with two-dimensional control and synchronise it to a near-counterpropagating ultrarelativistic electron bunch. 
Interaction at a finite angle produces forward-scattered radiation along an unimpeded path, facilitating its use in subsequent applications, while protecting the laser chain from damaging counterpropagating beams.
While a finite-angle geometry would limit the interaction length in a conventional focusing geometry, the flying focus can be matched to the electron-beam path, enabling extended laser-electron overlap.

Maintaining $a_0<1$ preserved linear electron motion, resulting in a higher spectral density and reduced divergence of the generated X-rays, compared to equivalent conventional focusing at high intensity without spatiotemporal control.
Even with the large energy spread of the electron bunch, this results in a more than 2-fold increase in spectral density for \unit[0.5]{MeV} photons incident on the detector.
Further benefits could be obtained with higher-quality electron bunches, such as those already demonstrated at BELLA HTU \cite{Barber2025PRL} and with similar laser systems such as the LUX beamline at DESY, Hamburg \cite{Kirchen2021PRL}.
With a high-quality \unit[250]{MeV} bunch (1.5\% energy spread and \unit[1]{mrad rms} divergence), the reduced spectral broadening of the flying-focus-enhanced Thomson scattering would result in a factor of three increase in the angularly integrated spectral density at \unit[1]{MeV}, compared to equivalent conventional focusing
(see \cref{fig:future_xray_spectrum}).
With the reduced divergence, this also corresponds to a 25-times increase in spectral brightness.
For comparison, the X-ray peak spectral brightness in our experiment is estimated as \unit[$\sim 3\times10^{20}$]{photons/mm$^2$/mrad$^2$/s/0.1\%BW} at \unit[$0.5$]{MeV}, assuming a \unit[30]{fs} electron bunch duration and a source size determined by the size of the scattering beam focal spot.
Further increases in yield and brightness could be achieved by scattering from a narrower electron bunch, for example by interacting closer to the accelerator exit, where {\textmu}m-scale bunch sizes have been demonstrated \cite{Weingartner2012PRSTAB}.

As flying-focus-enhanced Thomson scattering can be used to ensure efficient linear scattering, the X-ray yield increases linearly with the energy of the scattering pulse.
With a higher-energy laser system \cite{Danson2019HPLSE}, both the electron bunch energy and scattering laser energy can be increased and the advantages of flying-focus-enhanced Thomson (or inverse Compton) sources are even more apparent.
For example, using a \unit[10]{J} scattering laser and a \unit[1]{GeV} electron bunch, the spectral density can be increased by almost one order of magnitude, and the spectral brightness by more than two orders of magnitude at \unit[10]{MeV}, compared to equivalent focusing without spatiotemporal control, as shown in \cref{fig:future_xray_spectrum}.
This greater enhancement is due to the highly nonlinear nature of the interaction in the fully compressed case, which is mostly eliminated when using the flying focus.
The brightness achieved for a \unit[50]{pC} \unit[1]{GeV} electron bunch could be of the order $\unit[3\times10^{24}]{photons/mm^2/mrad^2/s/0.1\%BW}$, exceeding the performance of current all-optical inverse Compton sources \cite{Mirzaie2024NatPhoton}.


In conclusion, we have experimentally demonstrated a spatiotemporally engineered flying-focus pulse to enhance X-ray radiation from Thomson scattering.
The moving focal point of the pulse was programmed to match the velocity and trajectory of a relativistic electron bunch by precisely introducing longitudinal chromatic aberration, angular dispersion, and group delay dispersion.
This demonstration opens a new regime of structured-light control at high intensity through a comparatively simple but flexible setup that uses only standard refractive optics.
The interaction also shows that relativistic electron bunches can serve as probes of the evolving laser field, with the emitted X-rays verifying the programmed properties of the flying focus.
The results open the way to the next generation of flying-focus systems capable of sustaining programmable intensity structures, with implications for controlled laser-plasma acceleration and the development of bright, tunable radiation sources.
The enhanced X-ray sources enabled by this technique could significantly improve a range of applications, such as advanced blur-free imaging of high-Z materials through MeV radiography, isotope detection using nuclear resonance fluorescence \cite{Bertozzi2005NIMPB}, and studies of fundamental physics, such as photon-photon scattering \cite{Drebot_PRAB2017}.



\section*{Methods}

\subsection*{Electron beam generation and measurement}
\label{app:electrons}
The LWFA laser pulse was horizontally polarised and contained an energy of \unit[1.3]{J} in a pulse duration of \unit[32]{fs} (FWHM).
This was focused to a FWHM spot of size \unit[$19$]{{\textmu}m} $\times$ \unit[$17$]{{\textmu}m} (in the horizontal and vertical directions respectively), producing a vacuum peak intensity of \unit[$10^{19}$]{W cm$^{-2}$} (the rms variations in energy, spot size and intensity were within 2\%).
The focus was aligned to the entrance of a \unit[3]{mm} long conical gas jet containing 99.5\% helium and 0.5\% nitrogen, in which electron bunches were generated by ionisation injection and then accelerated via LWFA to relativistic energies.


The spectrum of the LWFA-generated electron bunches was measured using a magnetic spectrometer consisting of an adjustable electromagnet, here set to B = \unit[0.7]{T} with an integrated field strength $\int B\mathrm{d}x \approx \unit[0.3]{Tm}$, and charge-calibrated scintillating Lanex screens \cite{Nakamura2008RSI}. 
The spectrometer had a low-energy cut-off at \unit[25]{MeV}.

\begin{figure}
    \centering
    \includegraphics[width=8.45cm]{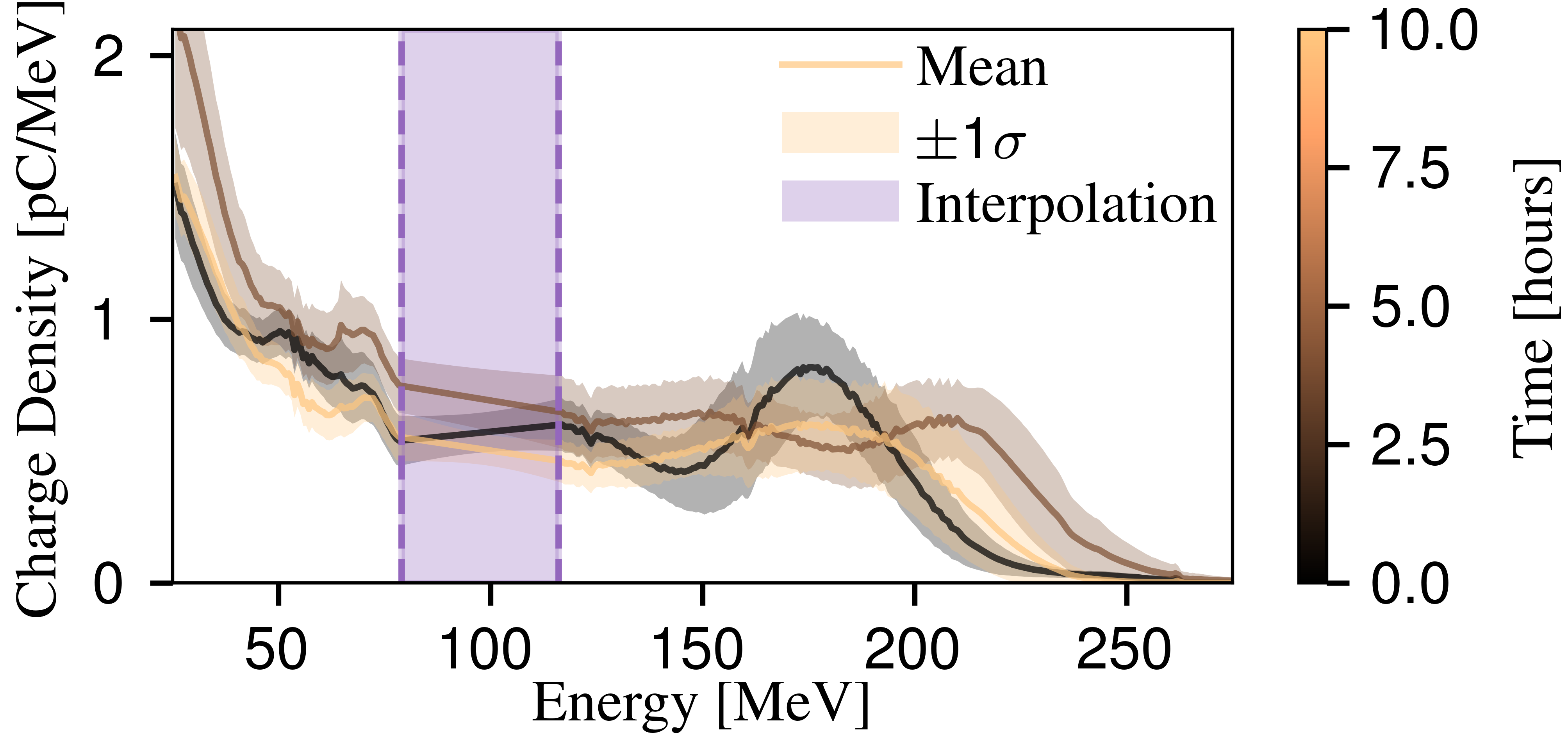}
    \caption{\textbf{Stable electron bunch performance over 10 hours of operation.}
    Average electron spectra (solid lines) and standard deviation (shaded bands), measured at the beginning of the shot day (grey), and after 4 (brown) and 10 hours (beige), respectively. The region between the dashed lines (purple shaded) indicates the region over which the spectrum was interpolated.
    }
    \label{fig:electron_spectrum}
\end{figure}

Due to imperfect overlap between the detector screens, the measured spectrum exhibits a \unit[$35$]{MeV} wide gap between \unit[80-115]{MeV} for the standard electromagnet setting (see \cref{fig:electron_spectrum}).
We measured this part of the spectrum by adjusting the strength of the electromagnet, and found that it can be approximated by a linear interpolation between the end points of the spectral gap.
The average contained charge was \unit[$14\pm2$]{\%} of the total bunch charge, while the error on the integrated charge resulting from the interpolation was up to \unit[$19\pm7$]{\%}.
This resulted in an overall negligible error on the total bunch charge of \unit[$3\pm1$]{\%}.
An additional Lanex screen was temporarily inserted before the magnet to measure the transverse profile and divergence of the electron bunch.

To reduce the bremsstrahlung background on the X-ray diagnostics, an additional permanent dipole magnet with $\int B\mathrm{d}x \approx \unit[0.1]{Tm}$ was inserted in the vacuum chamber during Thomson scattering shots.
As a result, the electron bunch was not measured simultaneously with the X-ray beam properties, but its performance was characterised intermittently throughout the experiment.

As shown in \cref{fig:electron_spectrum}, spectra recorded at the beginning of the shot day and after approximately \unit[4]{hours} and \unit[10]{hours} of operation were mutually consistent.
These measurements confirm that the electron spectrum remained stable over the duration of the experiment.

\subsection*{Flying-focus configuration and collision geometry}

The scattering laser was vertically polarised and contained an energy of \unit[0.5]{J}.
The interaction angle between the LWFA and the scattering laser was determined by imaging the emission from the plasma columns of each laser pulse separately from above.
In combination with the amount of LCA introduced by the focusing lens, this determined the amount of angular dispersion required to match the chromatic line focus of the scattering laser to the electron beam axis.

A ray-tracing model of the pulse compressor was used to calculate the effect of grating translations and rotations on the group delay dispersion and angular dispersion of the laser pulse. 
This determined the required grating motions to set the required group delay dispersion and angular dispersion correctly.
The spatial-spectral distribution of the flying focus was subsequently characterised using spectrally resolved imaging of the focal plane.
In order to set the zero-GDD position of the gratings, a frequency-resolved optical gating diagnostic (SHG-FROG Swamp Optics Grenouille) was used to find the grating position that flattened the spectral phase of the laser.
The GDD values were then determined from the relative change in grating separation from this position using the calibration from the numerical modelling.

\subsection*{X-ray generation and diagnostics}

After initial alignment of the scattering pulse to the LWFA drive pulse and synchronisation using a fast photodiode, fine alignment was achieved by scanning the scatter lens position and relative timing of the two laser pulses to maximise the Thomson X-ray signal.
The focal velocity was varied by scanning the grating separation, which added GDD to the scattering pulse.
The relative delay introduced by this motion was compensated with a separate delay stage to resynchronise to the electron bunch.

The X-rays were emitted in a narrow cone around the electron bunch axis and propagated through a \unit[2]{mm}-thick aluminium vacuum window at the end of a vacuum pipe to a suite of X-ray diagnostics outside the vacuum chamber at $\unit[4.65]{m}$ from the interaction point.
The aperture of the vacuum pipe limited the field of the detectors to $\approx\unit[11]{mrad}$.
The transverse profile and deposited energy were measured using a pixelated profiler consisting of scintillating thallium-doped caesium iodide (CsI:Tl) crystals, while the spectral shape was measured using a CsI:Tl spectrometer \cite{Behm_RSI2018}.


\subsection*{X-ray detector calibration and spectral retrieval}
\label{app:xrays}

\begin{figure}
    \centering
    \includegraphics[width=8.45cm]{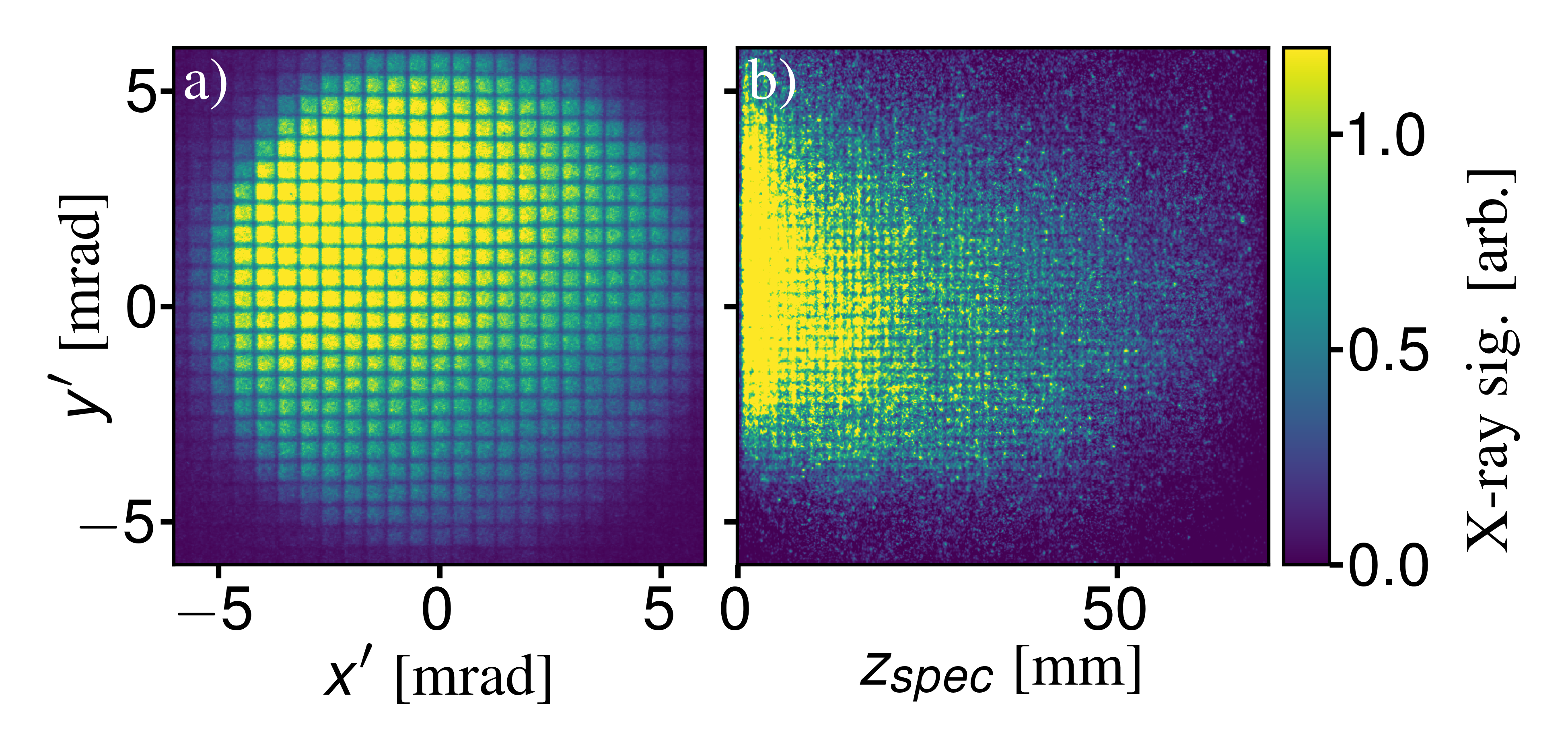}
    \caption{\textbf{Scintillator-based diagnostics measuring Thomson X-ray profile and spectrum.}
    \textbf{a)} X-ray profiler measures the transverse distribution of the X-ray beam truncated by the beam pipe. \textbf{b)} Spectrometer measures the penetration depth $z_{spec}$ of the X-ray beam entering from the left into the array of crystals, which can be used to retrieve the spectrum.
    }
    \label{fig:example_gammadiags}
\end{figure}

The X-ray profiler consisted of a $30\times30$ array of CsI:Tl crystals, each measuring $\unit[2]{mm} \times \unit[2]{mm} \times \unit[40]{mm}$, with the long dimension oriented parallel to the propagation path of the X-rays.
The spectrometer consisted of a $56\times56$ array of CsI:Tl crystals, each measuring $\unit[1]{mm} \times \unit[1]{mm} \times \unit[30]{mm}$, with the long dimension oriented perpendicular to the propagation path of the X-rays.
This orientation allowed the penetration depth of the incident radiation to be measured \cite{Behm_RSI2018}.
Both detectors were imaged using an intensified CCD camera (Andor iStar), with example images shown in \cref{fig:example_gammadiags}.

The diagnostics were calibrated using bremsstrahlung generated by inserting the electron beam profiler (Kodak Lanex Fast Front at 45 degrees) \cite{Cho_IEEE2008} into the beam path of the electrons.
The measured detector response was compared with Geant4 modelling to obtain a correction factor \cite{Behm_RSI2018}, accounting for crystal and imaging defects, and an absolute calibration of the deposited energy in the detector \cite{Gerstmayr_arxiv2025}.
The most significant error on the calibration was the shot-to-shot fluctuation of the electron bunch charge of $\approx 26\%$.

The parametrisation for the spectral shape that was used for the spectral retrieval was
\begin{equation}
    dN/dE = A (E/E_{crit})^{\mu} \exp[-(E/E_{crit})^{\nu}],
\end{equation}
which was found to enable a good fit for a wide range of parameters in simulations. 
We treat $E_{crit}$, $\mu$ and $\nu$ as free parameters to control the spectral shape to match the experimental signal using nonlinear least-squares fitting.
Using the fitted spectral shapes and the absolute calibration of the deposited energy, we can infer the total number of photons incident on the profiler.
This is under the assumption of constant spectral shape across the profiler.
The error on the calibration and the spectral shape combine to a systematic error on the photon yield of $\approx 35\%$.

This form extends the synchrotron-inspired spectrum used by Cole \textit{et al.} \cite{Cole2018PRX} (with fixed $\mu=-2/3$ and $\nu=1$) by allowing independent variation of the low-energy slope and high-energy cut-off. 
We use this expression as an empirical fitting approach to capture the simulated and retrieved spectral shapes, rather than as a test of a particular emission model.

The data were background subtracted using shots without the scattering beam.
The most significant source of background was bremsstrahlung, with betatron radiation and Compton photons emitted by electrons below the low-energy spectrometer cut-off being largely suppressed at the detector plane due to the aluminium vacuum window.

\subsection*{Numerical modelling of relative yield}
\label{app:numerical_model}

The laser field amplitude in the interaction region was calculated using a customised version of the Python code LASY \cite{LASY_code}.
The laser was initialised to match the experimental laser properties at the final focusing optic. The spatial profile was fitted as a super-Gaussian of order 3.58 and a $1/e^2$ intensity width of \unit[23.4]{mm}. The spectrum had a FWHM of \unit[21]{nm}, and total energy of \unit[0.5]{J}.
The longitudinal chromatic aberration was added as a frequency-dependent spatial phase term to match the calculated effect of the fused silica singlet lens. 
The numerical grid used 128 points along the transverse spatial axes covering \unit[93.64]{mm} and 512 points in the temporal axis covering \unit[11.4]{ps}.

The field at the nominal focal plane was calculated via fast Fourier transforms using the Fraunhofer approximation, with Fresnel diffraction used to propagate over the interaction region. The angular dispersion was calculated using a frequency dependent transverse shift, while the group delay dispersion was added by adding a frequency dependent phase term. For a given value of GDD and angular dispersion, the complex field was calculated over the $x,z$ plane at a given time $t$. The field was then sampled at a position corresponding to a test electron location at that instant. This was repeated for 200 time values, covering the full interaction time of the electron with the field to build up the field experienced by the electron during its transit. The total radiated energy was then estimated by integrating the classical Larmor power formula over this trajectory. 
The spatiotemporal offset of the electron compared to the field was selected to maximise the total radiated energy for each value of $\theta$ and GDD. To compare to the measured X-ray signals, the total energy was normalised and scaled to match the experimental measurements for \unit[$\beta<-5000$]{fs$^2$} for each value of $\theta$.
In finding the best-fitting flying-focus angle to the experimental data, Monte Carlo error propagation was used, assuming Gaussian errors in the measured X-ray signal and GDD values.

Ensembles of results were produced by randomly sampling from the probability distributions of each experimental measurement.
For each of the sets of such results, the flying-focus angle that produced the best fit was found.
The mean and $\pm1$ standard deviation angles of best fit from this ensemble were then calculated.
The model evaluated at the mean angle and the corresponding $\pm1$ standard deviation angles are plotted as the solid and dashed lines in \cref{fig:xray_vs_chirp}.

\bibliography{references}

\section*{Acknowledgements} \label{sec:acknowledgements}

We gratefully acknowledge the support of Zachary
Eisentraut, Mark Kirkpatrick, Federico Mazzini, Mackinley
Kath, Joe Riley, Arturo Magana, Teo Maldonado
Mancuso, Hai-En Tsai, Chetanya Jain, Nathan Ybarrolaza, Derrick McGrew, Carl Schroeder, Jens Osterhoff and Eric Esarey.
We also thank C.I. Hojbota and M.C. Downer for providing the X-ray spectrometer.
M.J.V.S. acknowledges support from the Royal Society URF-R1221874.
R.F. and A.G.R.T. acknowledge support from DOE NNSA Center of Excellence under Cooperative Agreement No. DE-NA0003869.
G.S. acknowledges support from EPSRC EP/V049186/1.
The work of M.V.-G., A.D.P., D.H.F., J.P.P., D.R., and H.G.R. is supported by the Office of Fusion Energy Sciences under Award Numbers DE-SC0021057, the Department of Energy National Nuclear Security Administration under Award Number DE-NA0004144, the University of Rochester, and the New York State Energy Research and Development Authority.
A.D.P. is partially supported by the U.S. National Science
Foundation Mid-scale Research Infrastructure Program
under Award No. PHY-2329970.
This work was supported by the U.S. Department of Energy Office of
Science Offices of Fusion Energy Sciences, High Energy Physics, and
LaserNetUS (lasernetus.org) under Contract No. DE-AC02-05CH11231 and DE-SC0021057.
We acknowledge the support of the Central Laser Facility Vulcan dark period community support programme 24-2.

\section*{Author Contributions} 

E.G. and M.J.V.S. conceived and designed the study.
E.G., C.M., R.F., M.V.-G., C.B., Q.C., A.J.G., B.G., R.J., A.L., A.M., K.N., L.O.-H., A.P., J.v.T. and M.J.V.S. developed the experimental setup, performed the experiment, and analysed the data.
E.G. and M.J.V.S. carried out the numerical modelling and wrote the manuscript.
All authors critically reviewed and revised the manuscript and approved the final version.

\section*{Competing Interests Statement} \label{sec:competing_interests}
The authors declare no competing interests.


\end{document}

%% file: authors.tex
\newcommand{\QUB}{School of Mathematics and Physics, Queen's University Belfast, BT7 1NN, Belfast UK}

\newcommand{\JAI}{The John Adams Institute for Accelerator Science, Imperial College London, London, SW7 2AZ, UK}

\newcommand{\CLF}{Central Laser Facility, STFC Rutherford Appleton Laboratory, Didcot OX11 0QX, UK}

\newcommand{\UMICH}{G\'erard Mourou Center for Ultrafast Optical Science, University of Michigan, Ann Arbor, MI 48109-2099, USA}

\newcommand{\DESY}{Deutsches Elektronen-Synchrotron DESY, Notkestr. 85, 22607 Hamburg, Germany}
\newcommand{\UH}{University of Hamburg, Department of Physics, Jungiusstr. 9, 20355 Hamburg, Germany}

\newcommand{\LLE}{Laboratory for Laser Energetics, University of Rochester, Rochester, NY, USA}

\newcommand{\BELLA}{Lawrence Berkeley National Laboratory, Berkeley, CA, 94720, USA}
\newcommand{\ROCHESTER}{Physics Department, University of Rochester, Rochester, NY, USA}
\newcommand{\ELIBL}{ELI Beamlines Facility, The Extreme Light Infrastructure ERIC, 252 41 Doln\'{i} B\v{r}e\v{z}any, Czech Republic}

\author{E.~Gerstmayr}
\email[Correspondence email address: ]{e.gerstmayr@qub.ac.uk}
\affiliation{\QUB}

\author{C.~Mariani}
\affiliation{\DESY}
\affiliation{\UH}

\author{R.~Fitzgarrald}
\affiliation{\UMICH}

\author{M.~VanDusen-Gross}
\affiliation{\LLE}

\author{C.~Berger}
\affiliation{\BELLA}

\author{Q.~Chen}
\affiliation{\BELLA}

\author{A.~Di Piazza}
\affiliation{\LLE}
\affiliation{\ROCHESTER}

\author{M.S.~Formanek}
\affiliation{\ELIBL}

\author{D.H.~Froula}
\affiliation{\LLE}

\author{C.G.R.~Geddes}
\affiliation{\BELLA}

\author{A.J.~Gonsalves}
\affiliation{\BELLA}

\author{B.~Greenwood}
\affiliation{\BELLA}

\author{R.~Jacob}
\affiliation{\BELLA}

\author{A.~Lu}
\affiliation{\BELLA}

\author{A.~McIlvenny}
\affiliation{\BELLA}

\author{K.~Nakamura}
\affiliation{\BELLA}

\author{L.~Obst-Huebl}
\affiliation{\BELLA}

\author{J.P.~Palastro}
\affiliation{\LLE}

\author{A.~Picksley}
\affiliation{\BELLA}


\author{K.~Poder}
\affiliation{\DESY}

\author{D.~Ramsey}
\affiliation{\LLE}

\author{H.G.~Rinderknecht}
\affiliation{\LLE}

\author{G.~Sarri}
\affiliation{\QUB}


\author{A.G.R.~Thomas}
\affiliation{\UMICH}

\author{J.~van Tilborg}
\affiliation{\BELLA}

\author{M.J.V.~Streeter}
\email[Correspondence email address: ]{m.streeter@qub.ac.uk}
\affiliation{\QUB}